\documentclass[journal]{IEEEtran}

\usepackage{graphicx}
\usepackage{amsmath,bbm,amssymb,amsfonts,amstext,amsopn,sansmath}
\usepackage{cite}
\usepackage{balance}
\usepackage{url}
\usepackage{epsfig}
\usepackage{setspace}
\usepackage{stmaryrd}
\usepackage{psfrag}	
\usepackage{multirow}
\usepackage{makecell}
\usepackage{float}
\usepackage[process=auto]{pstool}
\usepackage{etoolbox}
\usepackage{algorithm}
\usepackage{algorithmic}
\allowdisplaybreaks
\usepackage{textcomp}
\usepackage{flushend}

\usepackage{clipboard}
\newclipboard{Paper_V5}
\newcommand{\copyablelabel}[1]{\label{#1}} 

\usepackage{xcolor}
\makeatletter
\let\myorg@bibitem\bibitem
\def\bibitem#1#2\par{%
	\@ifundefined{bibitem@#1}{%
		\myorg@bibitem{#1}#2\par
	}{%
		\begingroup
		\color{\csname bibitem@#1\endcsname}%
		\myorg@bibitem{#1}#2\par
		\endgroup
	}%
}
\newcommand{\highlightref}[1]{\expandafter\newcommand\expandafter*\csname bibitem@#1\endcsname{blue}}
\makeatother

\newcommand{\x}{{\mathsf{x}}}
\newcommand{\y}{{\mathsf{y}}}
\newcommand{\z}{{\mathsf{z}}}

\newcommand{\jj}{\mathsf{j}}

\newcommand{\e}{\mathsf{e}}

\newcommand{\dd}{\mathrm{d}}

\def\bp{\mathbf{p}}

\def\bPsi{\boldsymbol{\Psi}}

\def\cP{\mathcal{P}}
\def\cA{\mathcal{A}}

\newtheorem{remk}{Remark}

\newtoggle{OneColumn}
\toggletrue{OneColumn}

\newtoggle{arXiv}
\togglefalse{arXiv}

\newcommand{\BLUE}{\color{black}}

\newcommand{\BLACK}{\color{black}}

\usepackage{lipsum}
\def\@IEEEBIOphotowidth{1cm}    
\def\@IEEEBIOphotodepth{1cm}   
\def\@IEEEBIOhangwidth{1.2cm}    
\def\@IEEEBIOhangdepth{1.2cm}    

\EndPreamble
\begin{document}
\title{Low-to-Zero-Overhead IRS Reconfiguration: Decoupling Illumination and Channel Estimation}

\author{Vahid Jamali,~\IEEEmembership{Member,~IEEE,} George C. Alexandropoulos,~\IEEEmembership{Senior~Member,~IEEE,}\\ Robert Schober,~\IEEEmembership{Fellow,~IEEE,} and H. Vincent Poor,~\IEEEmembership{Life~Fellow,~IEEE}  
	\thanks{V. Jamali and H. Vincent Poor  are with the Department of Electrical and Computer Engineering, Princeton University, Princeton, NJ 08544 USA (e-mail: \{jamali, poor\}@princeton.edu).}
	\thanks{G. C. Alexandropoulos is with the Department of Informatics and Telecommunications, National and Kapodistrian University of Athens, Panepistimiopolis Ilissia, 15784 Athens, Greece. (e-mail: alexandg@di.uoa.gr).}
\thanks{R. Schober is with the Institute for Digital Communications at Friedrich-Alexander University Erlangen-N\"urnberg (FAU), Erlangen, Germany (e-mail: robert.schober@fau.de).}\vspace{-0.85cm}
}

\maketitle

\begin{abstract}
Most algorithms developed for the optimization of Intelligent Reflecting Surfaces (IRSs) so far require knowledge of full Channel State Information (CSI). However, the resulting acquisition overhead constitutes a major bottleneck for the realization of IRS-assisted wireless systems in practice. In contrast, in this paper, focusing on downlink transmissions from a Base Station (BS) to a Mobile User (MU) that is located in a blockage region, we propose to optimize the IRS for illumination of the area centered around the MU. Hence, the proposed design requires the estimation of the MU's position and not the full CSI. For a given IRS phase-shift configuration, the end-to-end BS-IRS-MU channel can then be estimated using conventional channel estimation techniques. The IRS reconfiguration overhead for the proposed scheme depends on the MU mobility as well as on how wide the coverage of the IRS illumination is. Therefore, we develop a general IRS phase-shift design, which is valid for both the near- and far-field regimes and features a parameter for tuning the size of the illumination area. Moreover, we study a special case where the IRS illuminates the entire blockage area, which implies that the IRS phase shifts do not change over time leading to zero overhead for IRS~reconfiguration. 	
\end{abstract}

\begin{IEEEkeywords}
Intelligent reflecting surfaces, channel estimation, blockage area, illumination area, low-overhead design. 
\end{IEEEkeywords}
\vspace{-0.3cm}


\section{Introduction} 
Intelligent Reflecting Surfaces (IRSs) have attracted significant attention as an enabling technology for the realization of smart radio environments in future sixth generation (6G) wireless systems \cite{yu2021smart, RISE6G_2021}. However, the envisioned performance gain of IRSs mostly relies on the availability of Channel State Information (CSI), which given the typically large number of reflecting elements, denoted henceforth by $Q$, translates into a huge, often unaffordable, CSI acquisition overhead \cite{wei2021channel,najafi2020intelligent}.   

Various channel estimation techniques have been proposed in the literature; see \cite{wei2021channel} for a recent overview. For example, the ON/OFF protocol proposed in \cite{mishra2019channel} comprises $Q$ stages, where in each stage, only one reflecting element is ON and the corresponding cascaded Base Station (BS)-IRS-Mobile User (MU) channel is estimated. To improve the estimation accuracy, a Discrete Fourier Transform (DFT) based protocol was proposed in \cite{zheng2019intelligent} which again employs $Q$ stages but, in each stage, all IRS elements are ON and their reflection coefficients are designed based on one of the columns of the DFT matrix. In \cite{hu2021two}, the authors proposed an algorithm which estimates the BS-IRS and IRS-MU channels on two different time scales. In particular, they exploited the fact that the BS-IRS channel changes much more slowly than the IRS-MU channel, since the IRS and the BS are fixed nodes and the IRS is often deployed to ensure a Line-of-Sight (LoS) BS-IRS link. Nonetheless, the overhead of all of these channel estimation techniques scales with $Q$, which hinders their application for practically large~IRSs. 

Two categories of IRS channel estimation schemes have been proposed in the literature whose overhead does not scale with $Q$, namely sparsity- and codebook-based schemes \cite{wang2020compressed,jamali2021power}.  In particular, the sparsity of the wireless channel in the angular domain was exploited in \cite{wang2020compressed} to design a channel estimation algorithm. Therefore, the corresponding overhead scales with the number of dominant propagation paths of the wireless channel. In contrast,  in \cite{jamali2021power}, the authors proposed to estimate the end-to-end BS-IRS-MU channel only for a limited number of IRS phase-shift configurations drawn from a codebook. Since the channel estimation overhead scales with the codebook size, an IRS phase-shift profile was developed, which allows the design of small-size phase-shift codebooks. 

\begin{figure*}[t]
	\centering
	\includegraphics[width=0.55\textwidth]{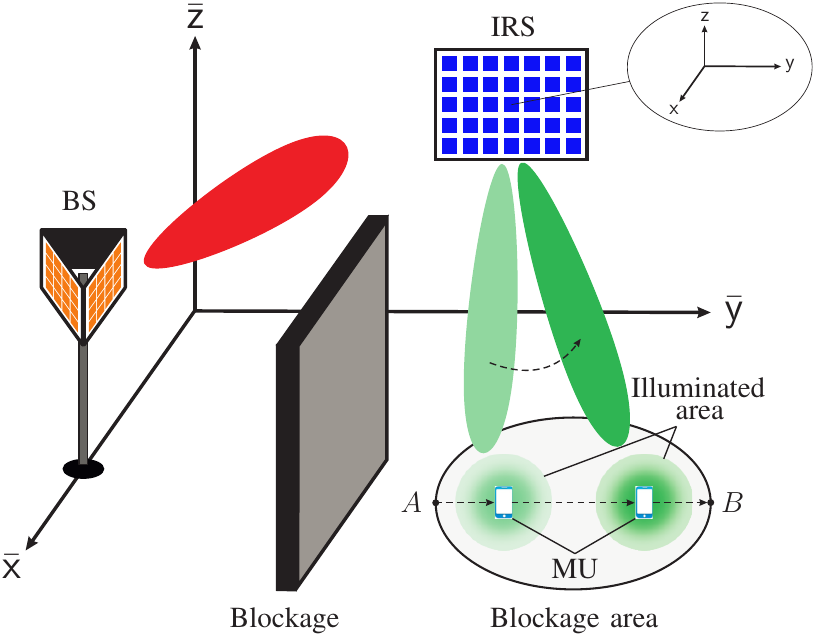}\vspace{-0.2cm}
	\caption{Schematic illustration of the considered IRS-assisted downlink communication system. Throughout the paper, we mainly use the coordinate system $\x\y\z$ and employ the coordinate system $\bar{\x}\bar{\y}\bar{\z}$ to only represent the positions of the multi-antenna BS, $Q$-element IRS, and single-antenna MU. The generic notations $\bp_i$ and $\bar{\bp}_i$ refer to points in the $\x\y\z$ and $\bar{\x}\bar{\y}\bar{\z}$ systems, respectively.
		\vspace{-0.5cm}}\label{fig:system_model}
\end{figure*} 

Regardless of which of the above CSI acquisition schemes is adopted, the common drawback of optimizing the IRS based on CSI is that, since channel gains change quickly, the IRS should be in principle frequently reconfigured (e.g., typical channel coherence times are on the order of milliseconds \cite{torres2021lower}). In contrast, in this paper, we propose to decouple the IRS reconfiguration from channel estimation. We focus on downlink transmissions from a BS to an MU that is located in a blockage region. The basic idea behind the proposed scheme is that the IRS is configured to illuminate the MU and is only reconfigured if the MU's Quality-of-Service (QoS) requirement cannot be met. Hence, the frequency of IRS reconfiguration does not explicitly depend on the channel coherence time, but on the mobility of the user, the size of the area illuminated by the IRS, and the MU QoS requirement. Once the IRS is configured, it acts as a virtual channel scatterer and the end-to-end BS-IRS-MU channel can be estimated based on conventional channel estimation techniques, and with a frequency that is dictated by the channel coherence time \cite{liu2018massive}. To facilitate the proposed approach, we develop a general IRS phase-shift design, which unlike those in \cite{jamali2021power,Laue2021irs} is valid for both the near- and far-field regimes and features a parameter for tuning the size of the illuminated area. We analyze the corresponding overhead and compare it with that of the existing schemes. Moreover, we study an interesting special case where the IRS illuminates the entire blockage area, avoiding the need for IRS reconfiguration. 



\section{Communication Setup}\label{Sec:System} 

\subsection{System  Model} 
\Copy{beamformer}{We consider a downlink system, where a multi-antenna BS wishes to serve a single-antenna  MU. We assume that there exists an area where direct coverage is not available due to, e.g., the presence of a blocking building, see Fig.~\ref{fig:system_model}. To address this issue, an IRS is deployed such that it has a LoS connection to both the BS and the MU, enabling the realization of a virtual non-LoS link between the BS and the MU. {\BLUE Since the BS and IRS are fixed nodes, we assume that the BS employs a fixed active beamformer directed towards the IRS.}} Hence, the equivalent baseband signal model is given by \cite{najafi2020intelligent}  
\begin{align}\copyablelabel{Eq:model_phase}
y = \sum_{q=1}^{Q} h_{r,q} \Gamma_{q} h_{i,q}s+ n,
\end{align}
where $s$, $y$, and $n$ denote the BS's data symbol, the MU's received signal, and the additive white Gaussian noise at the MU with zero mean and variance $\sigma_n^2$, respectively.  \Copy{effectivechannel}{{\BLUE Moreover, $h_{i,q}$ denotes the \textit{effective} channel coefficient (including the impact of BS beamforming) between the BS and the $q$-th reflecting element of the IRS}, and  $h_{r,q}$ denotes the channel coefficient between  the $q$-th reflecting element of the IRS and the single MU antenna.} 
Furthermore, $\Gamma_{q}\triangleq\bar{\Gamma}\e^{\jj \omega_q}$ is the reflection coefficient of the $q$-th IRS reflecting element, where $\bar{\Gamma}$ and $\omega_q$ are the amplitude and phase of $\Gamma_{q}$, respectively.

\vspace{-0.1cm}
\subsection{Scattering Integral}  
While the received signal model in \eqref{Eq:model_phase} is useful for characterizing the system on the small discrete-time scale, it is too general and abstract for an insightful design and performance analysis of the IRS. To cope with this issue, in this paper, we base our design on the LoS links, which for small sub-wavelength IRS element spacing, can be accurately characterized by the scattering integral \cite[Fig.~4]{najafi2020intelligent}. In particular, using the scalar representation of this integral, the reflected electric field at MU position $\bp_r$ can be derived as \cite{balanis2015antenna}
\begin{IEEEeqnarray}{lll}\label{eq:Er}
	E_r(\bp_r) &= \frac{1}{\jj\lambda}\int_{\y=-\frac{L_\y}{2}}^{\frac{L_\y}{2}} \int_{\z=-\frac{L_\z}{2}}^{\frac{L_\z}{2}}  
	E_i \e^{\jj\varphi(\bp)} \Gamma(\bp) 
	\frac{\e^{\jj \kappa \|\bp_r-\bp\|}}{\|\bp_r-\bp\|}
	\dd \y\dd \z, \,\quad
\end{IEEEeqnarray}
where $E_i$ and $\varphi(\bp)$ are the amplitude and phase, respectively, of the incident wave at point $\bp=[0,\,\y,\,\z]$ on the IRS; $\Gamma(\bp)=\tau\e^{\jj \omega(\bp)}$ is the ratio of reflected and incident electric fields (i.e., field reflection coefficient) where $\omega(\bp)$ is the phase shift applied by IRS at point $\bp$ and $\tau$ is a constant that ensures the passivity of the IRS, which in general depends on the incident and reflection angles \cite[Remark~1]{najafi2020intelligent}. $L_\y$ and $L_\z$ denote the IRS dimension along the $\y$- and $\z$-axes, respectively; $\lambda$ and $\kappa\triangleq\frac{2\pi}{\lambda}$ represent the wavelength and wave number, respectively. 

\Copy{transform}{{\BLUE For completeness, we show in \cite[Appendix]{jamali2021low}, which is an extended version of this letter, that the abstract model in \eqref{Eq:model_phase} is a consistent approximation of the physical model in \eqref{eq:Er} for a lossless IRS (i.e., $\bar{\Gamma}=1$), when assuming that $h_{r,q}$ and $h_{i,q}$ comprise only LoS links in an unobstructed propagation medium. For this case, the channel coeffitients are chosen as $h_{i,q}=\sqrt{D_{\rm tx}G_{\rm irs}} \lambda/(4\pi d_i) \e^{\jj\kappa d_{i,q}}$ and $h_{r,q}=\sqrt{G_{\rm irs} D_{\rm rx}}\lambda/(4\pi d_r) \e^{\jj \kappa d_{r,q}}$, respectively, where  $D_{\rm tx}$ and $D_{\rm rx}$ denote the directivity of the BS and MU antennas, respectively; $G_{\rm irs}=\frac{4\pi\tau A_{\rm uc}}{\lambda^2}$ denotes the effective power gain factor of each IRS element having area $A_{\rm uc}$; $d_i$ and $d_r$ are the distances between the BS and MU to the center of the IRS, respectively; and $d_{i,q}$ and $d_{r,q}$ denote the distances between the BS and MU to the $q$-th IRS unit cell, respectively.}}
		
Neglecting the noise, the transmit and received symbol powers $P_{\rm tx}\triangleq|s|^2$ and $P_{\rm rx}\triangleq|y|^2$, respectively, defined based on \eqref{Eq:model_phase}, can be obtained from \eqref{eq:Er} by relating the electric fields and power densities in space according to \cite[Appendix~A]{najafi2020intelligent}:
		\begin{align}  \copyablelabel{eq:relation}
			P_{\rm rx}=\frac{|E_r|^2}{2\eta}\frac{D_{\rm rx}\lambda^2}{4\pi},\,\,|E_i|^2=2\eta\frac{P_{\rm tx} D_{\rm tx}}{4\pi d_i^2}.
		\end{align}
where $\eta$ represents the free-space characteristic impedance.
 We next employ \eqref{Eq:model_phase} for channel estimation and \eqref{eq:Er} for IRS phase-shift design.

\subsection{User Mobility Model} 
Depending on the application of interest, various user mobility models have been proposed in the literature; see \cite{zheng2004recent} for an overview. In this paper, we adopt a simple user mobility model, which is a special case of the random waypoint mobility model \cite{zheng2004recent}. We specifically assume that the MU enters and leaves the area at random points $A$ and $B$, respectively, in Fig.~\ref{fig:system_model}, and crosses the area on a straight line with a fixed velocity $v$.

\section{Low-overhead IRS Reconfiguration} 


\subsection{Proposed Algorithm}

The proposed IRS reconfiguration and channel estimation algorithm consists of the following three sub-blocks:

\textbf{Sub-block 1 (MU Localization):} Recall that the IRS is deployed to have LoS connections to both the BS and the MU. Since the IRS and BS are fixed nodes, their relative positions can be estimated once, and then, considered as known. However, the position of the MU, denoted by $\bp_{ r}$, varies due to its mobility which can be estimated using existing localization algorithms (see, e.g., \cite{wymeersch2020radio,abu2021near}). Note that if the MU is in the far field of the IRS, it suffices for this localization sub-block to estimate the  Angle-of-Departure (AoD) from the IRS to the MU.  

\textbf{Sub-block 2 (IRS Phase-Shift Design):} We assume that the IRS phase-shift design is based only on the MU position or the required AoD in the case of far field. However, the IRS radiation pattern can be designed to be narrow \cite{najafi2020intelligent} or wide \cite{jamali2021power,Laue2021irs}, depending on the IRS configuration objective. This will be discussed in detail in the following Section~\ref{Sec:IRS}.

\textbf{Sub-block 3 (End-to-End Channel Estimation):} Once the IRS is configured, the BS and the MU treat the surface as a part of the end-to-end wireless channel $h_{\rm e2e}\triangleq\sum_{q=1}^{Q} h_{r,q}\Gamma_{q} h_{i,q} $, which can be estimated using standard (e.g., Least Squares (LS)) channel estimation techniques \cite{liu2018massive}. 

\begin{figure}[t]
	\centering
	\includegraphics[width=1\columnwidth]{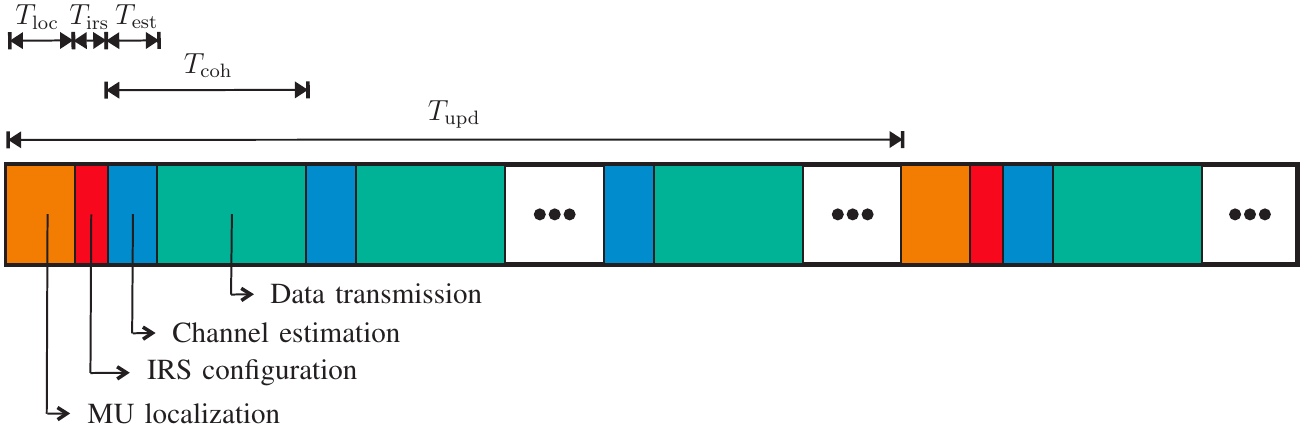}\vspace{-0.3cm}
	\caption{\Copy{CapFig2}{{\BLUE Block diagram of the proposed IRS-assisted downlink transmission scheme, including the sub-blocks of MU localization, IRS phase-shift design, and pilot-assisted end-to-end channel estimation. }} }\label{fig:schedule}	
\end{figure}

Figure~\ref{fig:schedule} illustrates how the above three sub-blocks are employed in the proposed communication protocol, where $T_{\rm loc}$ denotes the time duration needed by the adopted localization algorithm to estimate/update the MU's position; $T_{\rm irs}$ is the time duration needed to compute the IRS phase shifts and/or inform the IRS; $T_{\rm est}$ represents the time duration needed by the adopted channel estimation algorithm to estimate the end-to-end channel;  $T_{\rm coh}$ denotes the channel coherence time; and $T_{\rm upd}$ is the time duration between two consecutive configurations of the IRS. We assume that the IRS is updated (using sub-blocks 1 and 2) once the current IRS illumination pattern cannot support the MU's QoS any more. Thereby, we consider that the MU continuously calculates the average Signal-to-Noise Ratio (SNR) $\gamma$ (for which the fading is averaged out, and hence, it is determined by the LoS link), and once it falls below a certain threshold, denoted by $\gamma_{\rm thr}$, the MU sends a feedback message to the BS to initiate the MU localization and IRS  phase-shift update. The proposed IRS configuration and channel estimation algorithm is summarized in Algorithm~\ref{Alg:IRS}, where $\Delta$ is a parameter that controls the size of the illuminated area, as will be explained in the next~subsection.

{\BLACK
	\begin{algorithm}[t] 
		\caption{\small IRS Reconfiguration and Channel Estimation}\footnotesize
		\textbf{input:} Illumination parameter $\Delta$ and SNR requirement $\gamma_{\rm thr}$. \newline 
		\textbf{output:} Update of MU position $\bp_r$, IRS phase shifts $\omega_q$,~$\forall q$, and end-to-end channel $h_{\rm e2e}$.
		\begin{algorithmic}[1]\label{Alg:IRS}		
			\STATE Estimate/update $\bp_r$. \hfill \textit{$\%$ sub-block~1}
			\STATE Design  $\omega_q$,~$\forall q$, for given $\Delta$ and $\bp_r$.  \hfill \textit{$\%$ sub-block~2}
			\WHILE{$\gamma\geq\gamma_{\rm thr}$}		
			\STATE Estimate  $h_{\rm e2e}$ per channel coherence time. \hfill\textit{$\%$~sub-block~3}
			\item Update the average SNR $\gamma$ computation.
			\ENDWHILE 	
			\STATE Go to line 1. 		 
		\end{algorithmic}
	\end{algorithm}
}

      \begin{figure*}[t]
	\begin{minipage}{1\linewidth}
		\centering
		\includegraphics[width=0.75\textwidth]{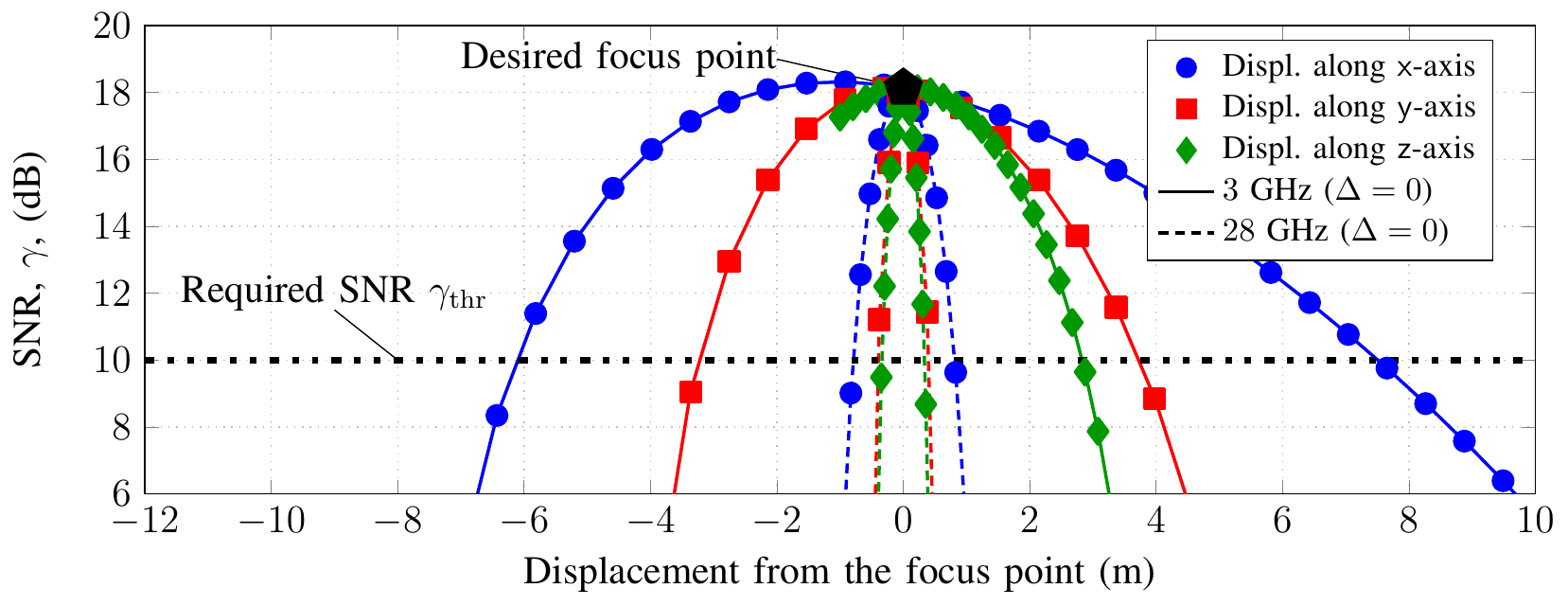}
	\end{minipage}
	\begin{minipage}{1\linewidth}
		\centering
		\includegraphics[width=0.75\textwidth]{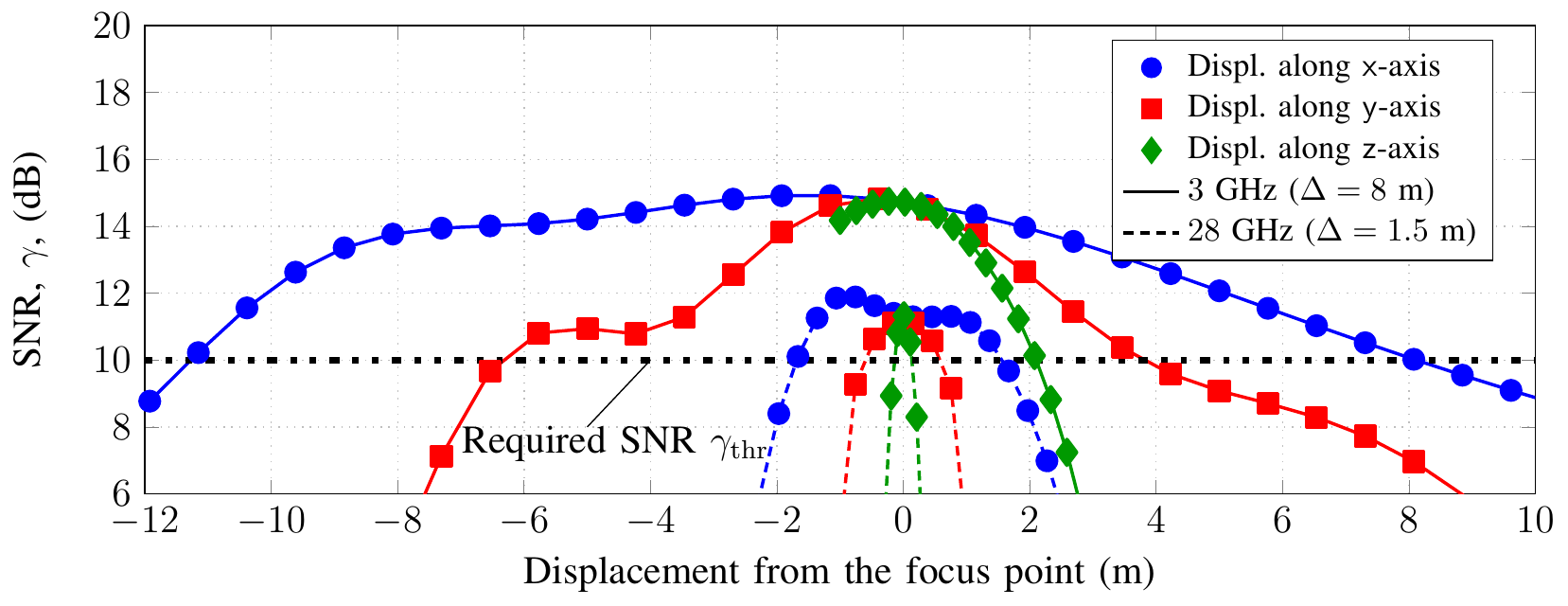}
	\end{minipage}
	\caption{Received SNR vs. displacement along the $\x$-, $\y$-, and $\z$-axes around the focus MU position $\bar{\bp}_{r}=[20,\, 60,\, 1]$~m for $L_\y=L_\z=50$~cm, IRS element spacing $d_\y=d_\z=0.5\lambda$, BS position $\bar{\bp}_{\rm bs}=[30,\, 0,\, 10]$~m, IRS central point position $\bar{\bp}_{\rm irs}=[0,\,50,\,5]$~m, $P_{\rm tx}=10$~dBm, $D_{\rm tx}=12$~dB,  $D_{\rm rx}=0$~dB, and $\sigma_n^2=WN_0N_{\rm f}$ with $N_0=-174$~dBm/Hz, $W=20$~MHz, and $N_{\rm f}=6$~dB. While the BS-IRS and IRS-MU distances are $d_i=58$~m and $d_r=22$~m, respectively, the far-field distance $d_{\rm F}=8(L_\y^2+L_\z^2)/\lambda$ is $40$~m and $373$~m for $3$~GHz and $28$~GHz, respectively.\vspace{-0.5cm}}\label{fig:snr_focus}
\end{figure*}

\vspace{-0.1cm}
\subsection{IRS Phase-Shift Design} \label{Sec:IRS}
\Copy{contribution}{{\BLUE Since various IRS-assisted localization schemes \cite{wymeersch2020radio,abu2021near} and conventional channel estimation schemes \cite{liu2018massive} have been proposed in the literature, in this section, we focus our attention on the design and analysis of efficient IRS phase-shift configurations enabling narrow and wide illuminations, respectively.}} 

 \subsubsection{Beam  Focusing} In order to maximize the power at the MU position $\bp_r$ (i.e., maximize $|E_r|$ via \eqref{eq:Er}), the following phase shift needs to be applied at point $\bp$ on the IRS:
 \begin{IEEEeqnarray}{lll}\label{Eq:w_focus}
 	\omega_{\rm F}(\bp|\bp_r)=-\kappa \|\bp_r-\bp\|-\varphi(\bp).
 \end{IEEEeqnarray}
Substituting the above IRS phase shifts in \eqref{eq:Er} and then in the $P_{\rm rx}$ expression in \eqref{eq:relation} yields the  maximum average SNR: 
 \begin{IEEEeqnarray}{lll}
 	\gamma_{\rm max}\triangleq\frac{P_{\rm rx,max}}{\sigma_n^2}=\frac{P_{\rm tx} D_{\rm tx}D_{\rm rx}}{\sigma_n^2}\left(\frac{\tau L_\y L_\z}{4\pi d_i d_r}\right)^2,
 \end{IEEEeqnarray}
where $d_r$ denotes the distance between the IRS center and the MU, and we used $\|\bp_r-\bp\|\approx d_r$ to approximate the amplitude $E_i$ in \eqref{eq:Er}. The main parameter that determines the overhead of the proposed design is how the SNR decays around the focus point, which is numerically evaluated in the top subfigure in Fig.~\ref{fig:snr_focus} for an example scenario. \Copy{TirsTcoh}{As can be observed from this figure, for this setup with the SNR requirement of $10$~dB, the coverage size along the $\x$- and $\y$-directions around the focus point are $14$~m and $7$~m ($1.6$~m and $0.8$~m), respectively, for $3$~GHz ($28$~GHz) carrier frequency. {\BLUE Hence, we can conclude that, for typical walking speeds of approximately $1$~m/s \cite{zheng2004recent}, the IRS update time $T_{\rm upd}$ is on the order of seconds, which is orders of magnitude larger than typical channel coherence times; those times are on the order of milliseconds~\cite{torres2021lower}.}}

\subsubsection{Wide Illumination} In order to further reduce the IRS reconfiguration overhead, we propose to widen the reflected beam around the MU's location to increase the area in which the SNR is above the required threshold $\gamma_{\rm thr}$. \Copy{wide}{Let $\cP_{\rm irs}$, $\cP_{\rm blk}$, and $\cP_{\rm ilm}\subseteq \cP_{\rm blk}$ denote the sets of points on the IRS, the blockage area, and the area that we wish to illuminate by the IRS, respectively. {\BLUE Suppose that if the IRS is set for full focusing based on \eqref{Eq:w_focus}, only a fraction of the targeted area $\cP_{\rm ilm}$ is illuminated. To illuminate the entire desired area, we partition the IRS into sub-surfaces and split the targeted illumination area into sub-regions, where each sub-surface on the IRS illuminates the center of one sub-region.  A general formulation of this problem is a mapping $\mathcal{M}$ from each point $\bp\in\cP_{\rm irs}$ on the IRS to  one of the sub-regions in the targeted illumination area. Let $\cP_{\rm ilm}^{\rm sub}\subseteq\cP_{\rm ilm}$ denote the set containing the centers of the sub-regions. Given this mapping, the corresponding IRS phase shift for wide illumination, denoted by $\omega_{\rm W}(\bp|\cP_{\rm ilm}^{\rm sub})$,~$\forall \bp\in\cP_{\rm irs}$, can be obtained as a function of $\omega_{\rm F}(\bp|\bp_r)$ in \eqref{Eq:w_focus}, as follows:
  \begin{align} \copyablelabel{Eq:WideBeam}
  	 \omega_{\rm W}(\bp|\cP_{\rm ilm}^{\rm sub})=\omega_{\rm F}(\bp|\mathcal{M}(\bp))-\kappa\|\mathcal{M}(\bp)-\bp_{\rm irs}\|.
  \end{align}
}}The term $\|\mathcal{M}(\bp)-\bp_{\rm irs}\|$ is the reference distance from each point $\mathcal{M}(\bp)$ to the IRS center, $\bp_{\rm irs}\triangleq[\x_i,\,\y_i,\,\z_i]$, whose contribution is removed from $\omega_{\rm F}(\bp|\mathcal{M}(\bp))$. Intuitively, even for focusing, the term $\kappa\|\mathcal{M}(\bp)-\bp_{\rm irs}\|$ only contributes to the signal's phase at the focus point and not to the corresponding power. Since we are mainly interested in the power at the desired observation points within $\mathcal{P}_{\rm ilm}$, and not in the phase, the contribution of the reference distances, i.e., $\kappa\|\mathcal{M}(\bp)-\bp_{\rm irs}\|$, are removed from the phase-shift design in \eqref{Eq:WideBeam}.

\Copy{mapping}{ {\BLUE Next, we discuss our proposed choices of the mapping function $\mathcal{M}(\bp)$ and the set $\cP_{\rm ilm}^{\rm sub}$. In principle, $\cP_{\rm ilm}^{\rm sub}$ can be a set of \textit{discrete} points $\bp_r\in\cP_{\rm ilm}$, but which and how many points to choose are non-trivial tasks. More importantly, for a given discrete set $\cP_{\rm ilm}^{\rm sub}$, the illumination patterns of the corresponding sub-regions interfere with each other and may cause severe fluctuations in the received powers particularly at the sub-region boundaries. To circumvent these challenges, we choose a \textit{continuous} set $\cP_{\rm ilm}^{\rm sub}$ spanning the entire targeted area, namely $\cP_{\rm ilm}^{\rm sub}\triangleq \cP_{\rm ilm}$. Therefore, $\mathcal{M}(\bp)\in\cP_{\rm ilm}$ becomes a continuous mapping from $\cP_{\rm irs}$ to $\cP_{\rm ilm}$, an example of which is provided in the following.}} 
   

\Copy{example}{   
   \textit{Example 1:} Let us assume  $L_\y=L_\z=L$ and a square area $\Delta\times \Delta$ for $\cP_{\rm ilm}$  parallel to the $\x\y$ plane  (i.e., the MU does not change its height). For this case, we propose the following simple mapping from $\cP_{\rm irs}$ to $\cP_{\rm ilm}$ (needed in \eqref{Eq:WideBeam}):
   \begin{align}\copyablelabel{Eq:MappSquare}
   	\mathcal{M}(\bp)\triangleq \left[\frac{\Delta}{L}\z+\x_i,\frac{\Delta}{L}\y+\y_i,\z_i\right],
   \end{align}
   which leads to the following IRS phase-shift design:
   \begin{align}\copyablelabel{Eq:PhaseWide} 
   	\omega_{\rm W}(\bp|\cP_{\rm ilm})  
   \!	=\! -\kappa \left(\|\mathcal{M}(\bp)\!-\!\bp\|\!\!-\!\!\|\mathcal{M}(\bp)\!-\!\bp_{\rm irs}\|\right)\!-\!\varphi(\bp).\!\!\!
   \end{align}
{\BLUE We can see from \eqref{Eq:MappSquare} that by varying point $\bp=(\y,\z)$ on the IRS, i.e.,   $(\y,\z)\in[-L/2,L/2]\times[-L/2,L/2]$,  the output of the mapping $[\tilde{\x},\tilde{\y},\z_i]\triangleq \mathcal{M}(\bp)$ varies and covers the entire targeted area, i.e., $(\tilde{\x},\tilde{\y})\in[\x_i-\Delta/2,\x_i+\Delta/2]\times[\y_i-\Delta/2,\y_i+\Delta/2]$. In other words, each part of the IRS locally focuses on one point in the set $\cP_{\rm ilm}$, and the phase shift across the IRS gradually varies such that all points in $\cP_{\rm ilm}$ are illuminated.}} The bottom figure of Fig.~\ref{fig:snr_focus} shows that for carrier frequencies of $3$~GHz and $28$~GHz  using $\Delta=8$~m and $1.5$~m, respectively, the areas supporting SNR threshold $\gamma_{\rm thr}=10$~dB along $(\x,\y)$-axis are respectively $(20,10)$~m and $(3,1.2)$~m, which implies a reduced IRS reconfiguration overhead compared to beam focusing.

\textit{Example 2:} Suppose that the BS and MU are in the far-field of the IRS. In this case, the phase shift $\omega_{\rm F}(\bp,\bp_r)$ given by \eqref{Eq:w_focus} reduces to the following linear phase shift \cite{najafi2020intelligent}:
\begin{IEEEeqnarray}{lll}  \label{Eq:Phaselinear}
	\omega_{\rm F}(\bp,\bPsi_r) = -\kappa\left[A_\y(\bPsi_i,\bPsi_r)\y+A_\z(\bPsi_i,\bPsi_r)\z\right],
\end{IEEEeqnarray}
where $\bPsi_i\triangleq(\theta_i,\phi_i)$ and $\bPsi_r\triangleq(\theta_r,\phi_r)$ denote respectively the incident and reflection angles with $\theta_u$ and $\phi_u$ for $u\in\{i,r\}$ being the elevation and azimuth angles in the IRS spherical coordinate system, respectively; $A_t(\bPsi_i,\bPsi_r)\triangleq A_t(\bPsi_i)+A_t(\bPsi_r)$ with $t\in\{\y,\z\}$; $A_\y(\bPsi_u)\triangleq \sin(\theta_u)\sin(\phi_u)$; and $A_\z(\bPsi_u)=\cos(\theta_u)$.
Note that in the far-field regime, the position $\bp_r\in\cP_{\rm blk}$ can be replaced by the corresponding AoD $\bPsi_r\in\varPsi_r$, where $\varPsi_r$ is the set of possible AoDs. For the ease of presentation, we assume that the range $\bPsi_r\in\varPsi_r$ translates into the following intervals for $\alpha_\y\triangleq A_\y(\bPsi_i,\bPsi_r)\in[A_\y^{\min},A_\y^{\max}]\triangleq\cA_\y$ and $\alpha_\z\triangleq A_\z(\bPsi_i,\bPsi_r)\in[A_\z^{\min},A_\z^{\max}]\triangleq\cA_\z$. Here, we propose the following mapping from $\bp\in\cP_{\rm irs}$ to $(\alpha_\y,\alpha_\z)\in\cA_\y\times\cA_\z\triangleq\cA_{\rm ilm}$ (or equivalently to $\bPsi_r\in\varPsi_r$):
\begin{IEEEeqnarray}{lll} \label{Eq:Maplinear}
	\mathcal{M}(\bp)\triangleq [a_\y\y+b_\y, \,a_\z\z+b_\z].
\end{IEEEeqnarray}
where $a_\y=(A_\y^{\max}-A_\y^{\min})/L$, $b_\y=(A_\y^{\max}+A_\y^{\min})/2$, $a_\z=(A_\z^{\max}-A_\z^{\min})/L$, and $b_\z=(A_\z^{\max}+A_\z^{\min})/2$. Substituting \eqref{Eq:Phaselinear} and \eqref{Eq:Maplinear} into the general IRS phase shift in \eqref{Eq:WideBeam}, we obtain the illumination of the desired width:   
\begin{IEEEeqnarray}{lll} 
	\omega_{\rm W}(\bp,\cA_{\rm ilm}) 
	= a_\y\y^2+b_\y\y +a_\z\z^2+b_\z\z.\quad\,\,
\end{IEEEeqnarray}
This phase-shift design is similar to the quadratic phase-shift profile proposed in \cite{jamali2021power} for the realization of small-sized IRS phase-shift~codebooks. 

\begin{remk}
\Copy{complexity}{{\BLUE The IRS phase shifts in \eqref{Eq:WideBeam} and \eqref{Eq:PhaseWide}  are  analytical, and hence, can be computed a priori for all possible $\cP_{\rm ilm}$, be stored in a memory, and then used on-demand for online communication.}}
\end{remk}

\begin{remk}
	\Copy{LocError}{\BLUE The designs in \eqref{Eq:WideBeam} and \eqref{Eq:PhaseWide} require the MU position, the estimation of which introduces an unavoidable error in practice \cite{wymeersch2020radio,abu2021near}. One way to account for such localization errors based on the designs in \eqref{Eq:WideBeam} and \eqref{Eq:PhaseWide} is to deliberately enlarge the size of the targeted illumination area~$\cP_{\rm ilm}$.}
\end{remk}

      \subsubsection{Full Illumination}
Next, we discuss an interesting special case where the IRS illuminates the entire blockage area and hence no IRS reconfiguration is  required, i.e., the IRS reconfiguration overhead is zero. This may be achieved in the phase-shift design proposed in \eqref{Eq:PhaseWide} by setting $\cP_{\rm ilm} =\cP_{\rm blk}$. In the following, we study for which parameter values, the MU's QoS can be met based on full illumination for an idealized scenario.  


\Copy{upper}{Let us assume that the blockage area is larger than what can be covered by focusing, as realized with \eqref{Eq:w_focus}. We assume an idealized IRS illumination where the entire power received by the IRS is uniformly distributed across the  blockage area. Note that such illumination cannot be realized by an IRS \cite{najafi2020intelligent}, but it provides a performance upper bound for the proposed IRS phase-shift design. Using \eqref{eq:relation}, this leads to the following SNR across the blockage area:
   \begin{align}\copyablelabel{eq:SNRuniform}
   	\gamma = \frac{P_{\rm tx} D_{\rm tx}D_{\rm rx}}{\sigma_n^2}
		\left(\frac{\lambda}{4\pi d_i}\right)^2
		\frac{L_\y L_\z A_\x(\bPsi_i)}{A_{\rm blk}},
   \end{align}
where $A_\x(\bPsi_i)\triangleq\sin(\theta_i)\cos(\phi_i)$ and $A_{\rm blk}$ denotes the size of the blockage area. {\BLUE The above expression can be used to characterize the tradeoff between the BS's transmit power, the IRS size, and the size of the blockage area where the SNR constraint $\gamma\geq\gamma_{\rm thr}$ holds across the entire illuminated~area.}}

  \begin{table}[t]
	\caption{Comparison of the overhead of CSI acquisition and IRS reconfiguration schemes.\vspace{-0.15cm}}\small
	\label{table:overhead} 
	\centering
	\scalebox{0.85}{
		\begin{tabular}{||l|c||}\hline
			\textbf{Schemes} & \textbf{Overhead} ($\triangleq\min\{\alpha,1\}$) \\\hline
			\makecell[l]{ON/OFF-based \cite{mishra2019channel} \\or DFT matrix-based \cite{zheng2019intelligent}} &  $\alpha=\frac{QN_{\rm plt} T_{\rm sym}}{T_{\rm coh}}$\\\hline
			\makecell[l]{Joint sparsity-based \cite{wang2020compressed} \\ and two-time scale-based \cite{hu2021two}} &  $\alpha = \frac{CN_{\rm pth} \log(N_{\rm grd}) T_{\rm sym}}{T_{\rm coh}}$\\\hline
			Codebook-based \cite{jamali2021power} &  $\alpha = \frac{N_{\rm cbk} N_{\rm plt}  T_{\rm sym}}{T_{\rm coh}}$\\\hline
			\makecell[l]{Prop. decoupling of illumination \\ and channel estimation} &  $\BLUE\alpha = \frac{T_{\rm loc}}{T_{\rm upd}}+\frac{(T_{\rm upd}-T_{\rm loc})N_{\rm plt} T_{\rm sym}}{T_{\rm upd}T_{\rm coh}}$\\\hline
		\end{tabular}
	}
\end{table}   

   \subsection{Overhead Comparison} 
\Copy{overhead}{{\BLUE Table~\ref{table:overhead} presents an approximate characterization of the overhead of the proposed IRS reconfiguration and CSI acquisition scheme, and that of several benchmark schemes from the literature for the considered communication system detailed in Section~\ref{Sec:System}, in terms of the underlying key parameters. We consider pilot symbols used only for channel estimation and localization, and neglect the time consumed for the feedback of the estimated CSI, MU position, and IRS phase shifts.}
	
\begin{figure*}[t]
	\centering
	\includegraphics[width=0.75\textwidth]{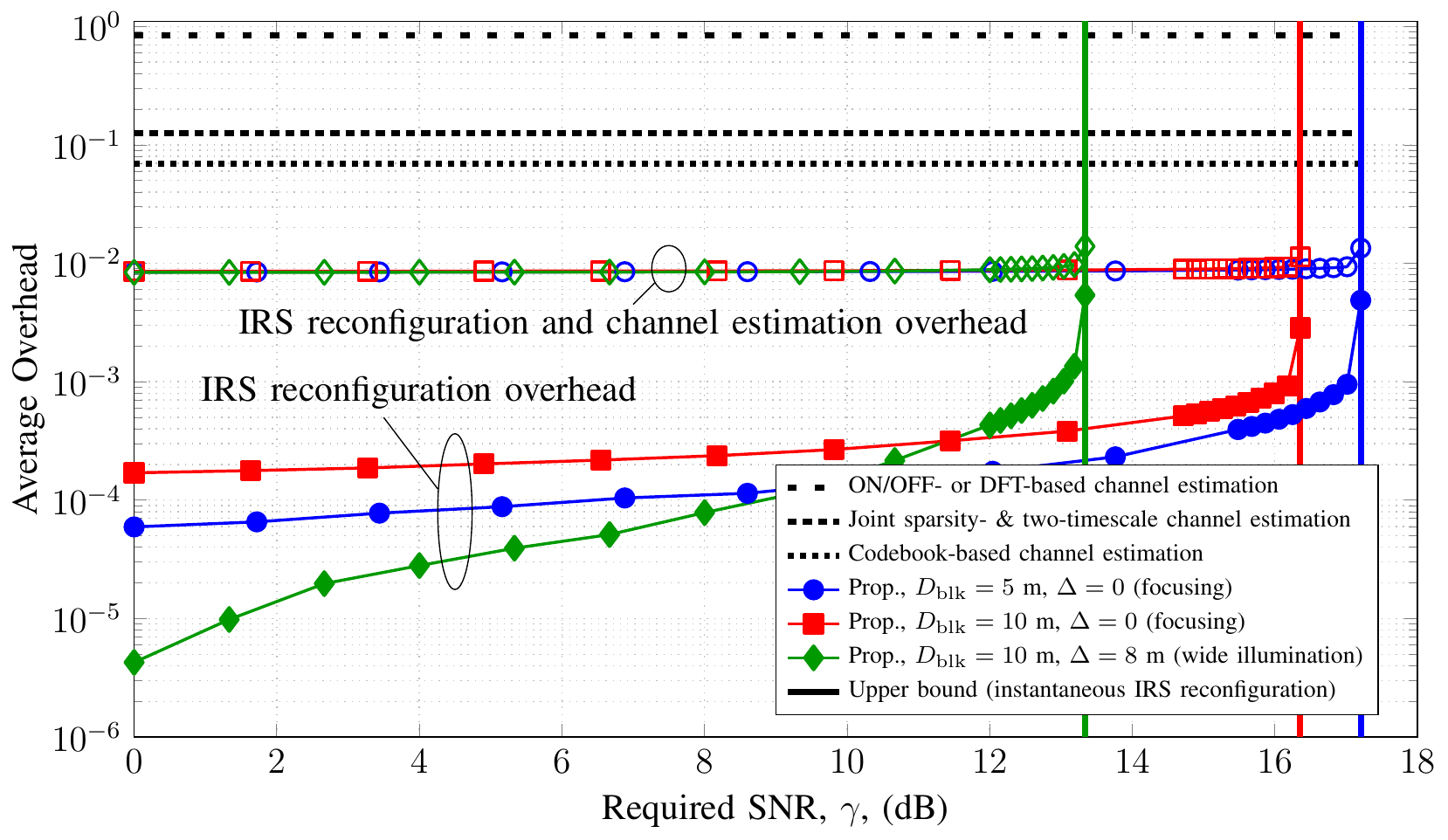}\vspace{-0.3cm}
	\caption{\Copy{FigSim1}{Average overhead vs. the required SNR  for the system parameters used in Fig.~\ref{fig:snr_focus} and $\bar{\bp}_{\rm blk}=[20,\, 60,\, 1]$, $v=0.75$~m/s, $N_{\rm plt}=3$, $N_{\rm pth}=5$, $N_{\rm grd}=20$, $N_{\rm cdb}=25$, $C=1$, $W_{\rm sub}=\frac{1}{T_{\rm sym}}=15$~kHz, and carrier frequency $3$~GHz (i.e., $Q=100$).}  \vspace{-0.3cm}}\label{fig:OverheadSNR}
\end{figure*} 
   
For the ON/OFF- and DFT matrix-based methods \cite{mishra2019channel,zheng2019intelligent}, we assume that in each of the $Q$ stages, $N_{\rm plt}$ pilot symbols of length $T_{\rm sym}$ are sent. We use the sparsity-based method in \cite{wang2020compressed} jointly with the two timescale-based method  in \cite{hu2021two}, where the overhead of BS-IRS channel estimation is neglected and the number of pilot symbols needed to estimate the IRS-MU channel  is assumed to be $CN_{\rm pth} \log(N_{\rm grd}) $ with $N_{\rm pth}$ and $N_{\rm grd}$ being the number of paths in the IRS-MU link and the number of discrete grid  points, respectively, and $C$ is a constant that depends on the adopted compressed sensing algorithm. The codebook size for codebook-based channel estimation is denoted by $N_{\rm cbk}$. {\BLUE For the proposed scheme, the value of $T_{\rm loc}$ in general depends on the adopted localization scheme. However, if sparsity-based localization is adopted and the MU's position is assumed to be along the dominant path, the number of required pilot symbols, and hence, $T_{\rm loc}$ is upper bounded by $CN_{\rm pth} \log(N_{\rm grd})$  \cite{wang2020compressed}.} }

  \section{Simulation Results}\label{Sec:Sim} 
The communication setup considered in this section is schematically illustrated in Fig.~\ref{fig:system_model} and the values of the system parameters are identical to those used for Fig.~\ref{fig:snr_focus} unless otherwise stated. \Copy{Tcoh}{{\BLUE The channel coherence time is computed based on \cite[Eq. (8)]{torres2021lower} and is approximately $24$ ms at $3$~GHz and $v=0.75$~m/s.}} The blockage area is a circle with center $\bar{\bp}_{\rm blk}$ and diameter $D_{\rm blk}$. The illumination scheme in \eqref{Eq:PhaseWide} is adopted which reduces to the near-field focusing in \eqref{Eq:w_focus} for $\Delta=0$. 

\begin{figure*}[t]
	\centering
	\includegraphics[width=0.75\textwidth]{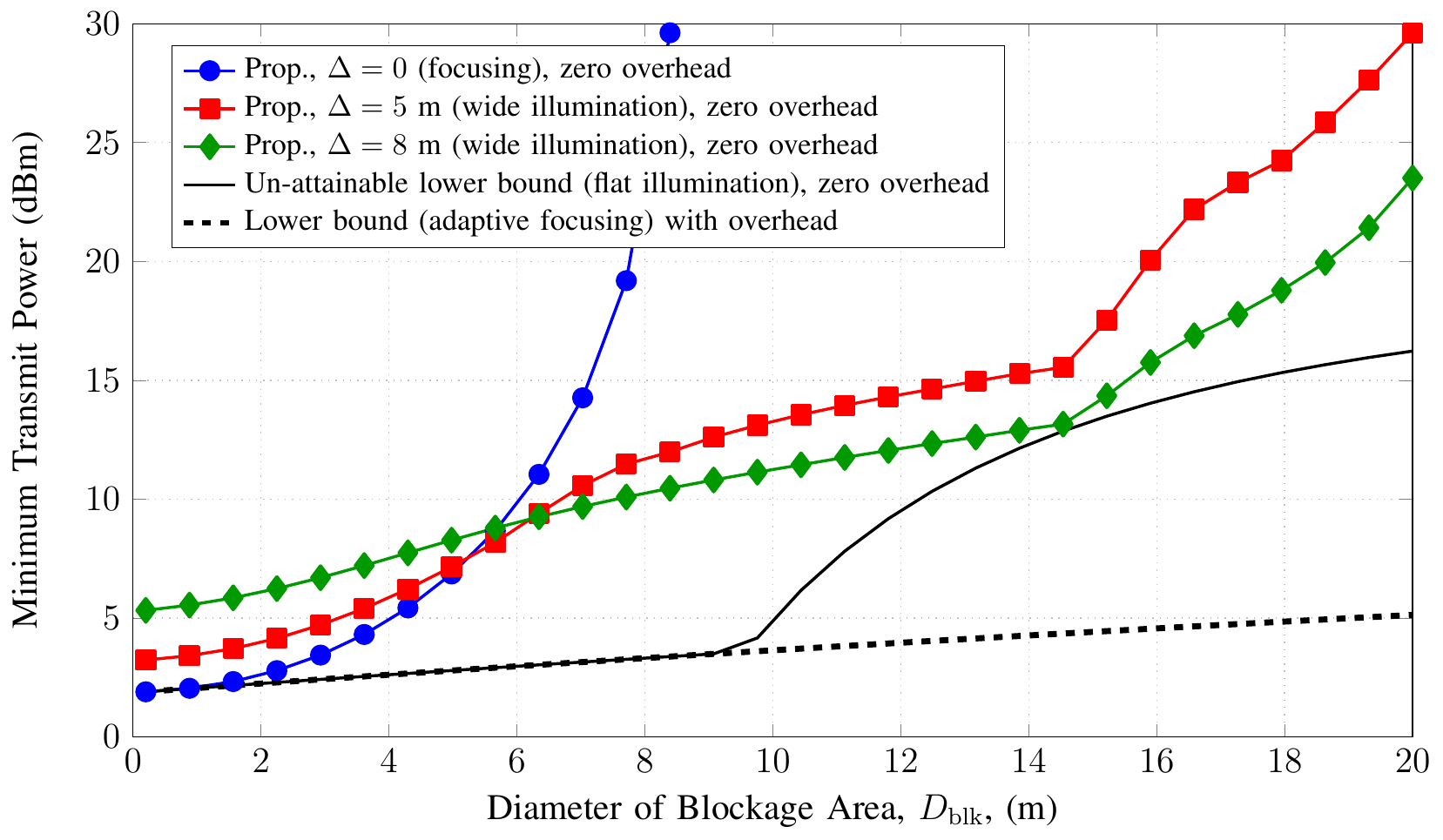}\vspace{-0.3cm}
	\caption{\Copy{FigSim2}{Minimum transmit power vs. the diameter of the blockage area for the system parameters used in Fig.~\ref{fig:snr_focus} and $\bar{\bp}_{\rm blk}=[20,\, 60,\, 1]$, required SNR $10$~dB, and carrier frequency $3$~GHz (i.e., $Q=100$).}  \vspace{-0.5cm}}\label{fig:PowerDr}
\end{figure*}

\Copy{Sim}{
{\BLUE In Fig.~\ref{fig:OverheadSNR}, we plot the overhead of only the IRS reconfiguration and the combined overhead of the IRS reconfiguration and channel estimation for the proposed scheme as a function of the required SNR.} Moreover, we include the channel estimation overhead  for the different benchmark schemes given in Table~\ref{table:overhead} for comparison. The vertical lines in Fig.~\ref{fig:OverheadSNR} represent the maximum SNR that is achievable for all points within the blockage area and can be attained only via instantaneous IRS reconfiguration. {\BLUE This figure suggests that, unless the required SNR is extremely close the maximum achievable SNR, the IRS reconfiguration overhead is orders of magnitude less than the channel estimation overhead. In other words,  at the cost of a negligible overhead,  the proposed scheme can support SNRs which are quite close (e.g., within $1$~dB) to the upper bound. Moreover, we observe from Fig.~\ref{fig:OverheadSNR} that, compared to near-field focusing ($\Delta=0$), wide illumination ($\Delta=8$~m) yields a lower IRS reconfiguration overhead for low required~SNR~values.}

In Fig.~\ref{fig:PowerDr}, we plot the minimum transmit power needed to support $\gamma_{\rm thr}=10$~dB via full illumination (i.e., zero IRS reconfiguration overhead) as a function of the blockage area diameter. As can be seen from this figure, for near-field focusing, the transmit power has to significantly increase as the blockage area size increases. {\BLUE In contrast, wider illumination requires a smaller transmit power for larger blockage areas.  An un-attainable lower bound for the transmit power  obtained from \eqref{eq:SNRuniform} and a lower bound  achievable by instantaneous focusing (incurring overhead) are also plotted for comparison.}
}


\section{\BLUE Conclusions}

\Copy{conclusion}{{\BLUE In this paper, we have shown that the overhead of IRS reconfiguration can be significantly reduced by decoupling it from channel estimation. Thereby, the IRS reconfiguration overhead  scales neither with the number of IRS elements nor with the channel coherence time, but it is a function of the MU's speed, its QoS requirements, and the IRS phase-shift design. Moreover, we have proposed a novel IRS phase-shift design which is valid for both near- and far-field transmission and features a parameter that tunes the width of the illuminated area. An interesting topic for future research is the use of machine learning tools for the localization and prediction of the IRS phase-shift configurations needed in the proposed scheme.}}

\bibliographystyle{IEEEtran}
\bibliography{References}


\appendix

In this section, we identify the conditions that relate the channel models in \eqref{Eq:model_phase} and \eqref{eq:Er}.  The analysis method is borrowed from \cite{balanis2015antenna} and is based on transforming the wave electrical field to the spatial power density using the Poynting  theorem, and then, transforming the field power density to the power absorbed/radiated  by the receive/transmit antenna given the antenna aperture.   

First, we note that the scattering integral in \eqref{eq:Er} assumes continuous surface approximation, whereas \eqref{Eq:model_phase} assumes a discrete array model. Therefore, we begin with approximating the integral in \eqref{eq:Er} by summation where the $\y$ and $\z$ axes are discretized with step sizes $d_\y$ and $d_\z$, respectively, i.e.,
\begin{IEEEeqnarray}{lll}\label{eq:Er_app}
	E_r(\bp_r) &= \frac{\tau}{\jj\lambda}\int_{\y=-\frac{L_\y}{2}}^{\frac{L_\y}{2}} \int_{\z=-\frac{L_\z}{2}}^{\frac{L_\z}{2}} \!\!\!\!  E_i \e^{\jj\varphi(\bp)}\frac{\e^{\jj \kappa \|\bp_r-\bp\|}}{\|\bp_r-\bp\|}\e^{\jj \omega(\bp)}\dd \y\dd \z, \,\quad\nonumber\\
	& \overset{(a)}{\approx} \frac{\tau}{\jj\lambda} \sum_{q=1}^{Q}
	E_i \e^{\jj\varphi(\bp_{q})}\frac{\e^{\jj \kappa \|\bp_r-\bp_q\|}}{\|\bp_r-\bp_q\|}\e^{\jj \omega(\bp_q)}d_\y d_\z\quad\nonumber\\
	&\overset{(b)}{\approx} \frac{\tau}{\jj\lambda} \sum_{q=1}^{Q}
	E_i \e^{\jj\varphi_q}\frac{\e^{\jj \kappa d_{r,q}}}{d_r}\e^{\jj \tilde{\omega}_q}d_\y d_\z,
\end{IEEEeqnarray}
where approximation $(a)$ is due to discretization with  $\bp_{q}$ denoting the $q$-th discrete point on the surface. Moreover,  approximation $(b)$ is due to the fact that, while  we consider the exact distance $d_{r,q}\triangleq \|\bp_r-\bp_q\|$ to compute the phase term $\e^{\jj \kappa d_{r,q}}$, we assume that the variation of the distance $d_{r,q}$ can be ignored for the amplitude term $\frac{1}{d_{r,q}}\approx\frac{1}{d_r}$, which is a widely-adopted approximation in antenna theory \cite{balanis2015antenna}.  Here, $d_r$ is the distance from the center of the IRS to the observation point $\bp_r$. For $(b)$, we further define $\varphi_q\triangleq \varphi(\bp_{q})$ and $\tilde{\omega}_q\triangleq\omega(\bp_q)$ for simplicity of presentation. 

The radiation power intensity (in Watt/m$^2$) at the IRS is
obtained as $S_i = \frac{P_{\rm tx}D_{\rm tx}}{4\pi d_i^2}$, where $P_{\rm tx}\triangleq |s|^2$ is the transmit power, $D_{\rm tx}$ denotes the directivity of the BS antenna, and $d_i$ is the distance between the BS and the center of the IRS. Moreover, for plane
waves\footnote{While the impinging wave may not be a plane wave across the entire IRS, the assumption of a plane wave is locally valid for the computation of the local electric fields on the IRS \cite{balanis2015antenna}.}, $S_i$ is related to the scalar electric field $E_i$ using the Poynting theorem as $S_i = \frac{|E_i|^2}{2\eta}$, where $\eta$ is the free-space characteristic impedance. Hence, assuming un-obstructed spherical propagation, the impinging electric field can be obtained as follows  
\begin{IEEEeqnarray}{lll}\label{eq:Ei_app}
	E_i \e^{\jj\varphi_q} = \sqrt{\frac{2\eta D_{\rm tx}}{4\pi d_i^2}} \e^{\jj\kappa d_{i,q}} s.
\end{IEEEeqnarray}
Similarly, the power collected by the receiver can be obtained as $P_{\rm rx} = S_r A_{\rm rx}$, where $S_r = \frac{|E_r|^2}{2\eta}$
is the radiation power intensity (Watt/m$^2$) at the MU and
$A_{\rm rx} = \frac{D_{\rm rx}\lambda^2}{4\pi}$ is the effective area of the MU antenna. For spherical propagation, the corresponding electric signal generated at the MU is given by
\begin{IEEEeqnarray}{lll}\label{eq:Yr_app}
	y = \sqrt{\frac{D_{\rm rx}\lambda^2}{2\eta 4\pi}} E_r+n.
\end{IEEEeqnarray}

Substituting \eqref{eq:Er_app} and \eqref{eq:Ei_app} into \eqref{eq:Yr_app} and simplifying the results yields
\begin{IEEEeqnarray}{lll}\label{eq:EiEr_app}
	y
	&= \sqrt{\frac{D_{\rm rx}\lambda^2}{2\eta 4\pi}} \times \frac{\tau}{\jj\lambda} \sum_{q=1}^{Q}
	\sqrt{\frac{2\eta D_{\rm tx}}{4\pi d_i^2}} \e^{\jj\kappa d_{i,q}} s\times \frac{\e^{\jj \kappa d_{r,q}}}{d_r}\e^{\jj \omega_q}d_\y d_\z+n \nonumber \\
	&=\!\!\sum_{q=1}^{Q} \! \underset{h_{r,q}}{\underbrace{\sqrt{\frac{G_{\rm irs} D_{\rm rx}\lambda^2}{(4\pi d_r)^2}} \e^{\jj \kappa d_{r,q}} }}
	\underset{\Gamma_{q}}{\underbrace{\e^{\jj 
				(\tilde{\omega}_q-\frac{\pi}{2})}}}
	\underset{h_{i,q}}{\underbrace{ \sqrt{\frac{D_{\rm tx}G_{\rm irs} \lambda^2}{(4\pi d_i)^2}} \e^{\jj\kappa d_{i,q}} }}  s \! +\! n.\quad\,\,\,\,
\end{IEEEeqnarray}
where $G_{\rm irs}=\frac{4\pi\tau A_{\rm uc}}{\lambda^2}$  is referred to as the effective power gain factor of each IRS element, where $A_{\rm uc}=d_\y d_\z$.
Therefore, under the assumption of unobstructed propagation, \eqref{Eq:model_phase} is consistent with \eqref{eq:Er} (up to the relaxation of the continuous surface approximation in \eqref{eq:Er})  if the BS-IRS and IRS-MU channel coefficients are chosen as $h_{i,q}=\sqrt{D_{\rm tx}G_{\rm irs}} \lambda/(4\pi d_i) \e^{\jj\kappa d_{i,q}}$ and $h_{r,q}=\sqrt{G_{\rm irs} D_{\rm rx}}\lambda/(4\pi d_r) \e^{\jj \kappa d_{r,q}}$, respectively, and the reflection coefficients are chosen as $\Gamma_{q}=\bar{\Gamma}\e^{\jj (\omega_q-\frac{\pi}{2})}$ with $\bar{\Gamma}=1$. Note that $\bar{\Gamma}=1$ implies a lossless IRS and the phase offset $-\frac{\pi}{2}$ originates from factor $1/\jj$ in \eqref{eq:Er}. 

Finally, we emphasize that while accounting for mutual coupling for discrete IRSs with sub-wavelength elements spacing is crucial in an array-based model \cite{ivrlavc2010toward}, the scattering integral in \eqref{eq:Er} inherently accounts for mutual coupling for the asymptotic case of a continuous IRS \cite{di2020smart}. However, we do not claim that under the above consistency conditions, the  array-based model in \eqref{Eq:model_phase} accounts for mutual coupling but only that the model in \eqref{Eq:model_phase}  becomes consistent with \eqref{eq:Er} in the asymptotic regime where $d_\y,d_\z\to 0$. Therefore, in general, the \textit{effective element factor} $G_{\rm irs}$ should not be considered the same quantity as the antenna directivity/gain (despite their similar forms), but merely a parameter that ensures the  consistency of \eqref{Eq:model_phase}  and \eqref{eq:Er}  in the asymptotic limit of small element spacing $d_\y,d_\z\to 0$.


\if{0}
\vspace{1cm}
\textbf{Possible extensions:}
\begin{itemize}
	\item Analysis of the beam pattern of focusing 
	\item Optimization-based design of near-field wide illumination
	\item Simulation of the end-to-end system, including localization, IRS reconfiguration, and channel estimation  
	\item Performance evaluation in terms of fading/channel scatterers 
	\item Explicit study of near- and far-field
	\item Study of the regime where satisfying the required SNR may not be infeasible
	\item Comparison with near-field broadcasting based on geometric optics
	\item Evaluation of overhead as a function of $Q$
\end{itemize}
\fi 

\end{document}